\documentclass[twocolumn,showpacs,preprintnumbers,amsmath,amssymb]{revtex4}
\usepackage{graphicx}
\usepackage{epsfig}
\usepackage{dcolumn}
\usepackage{bm}

\begin{document} 
\title{Statistics and Characteristics of Spatio-Temporally \\
Rare Intense Events in Complex Ginzburg-Landau Models}
\author{Jong-Won Kim$^1$ and Edward Ott$^{1,2}$}
\affiliation{$^1$Department of Physics, and Institute for Research in 
Electronics and Applied Physics, 
University of Maryland, College Park, Maryland  20742 \\ 
$^2$Department of Electrical and Computer Engineering,
University of Maryland, College Park, Maryland  20742} 
\date{\today} 
 
\begin{abstract}
We study the statistics and characteristics of rare intense events in two
types of two
dimensional Complex Ginzburg-Landau (CGL) equation based models. Our numerical 
simulations show finite amplitude collapse-like solutions which approach the
infinite amplitude solutions of the nonlinear Schr\"{o}dinger (NLS) equation 
in an appropriate parameter regime. We also determine the probability 
distribution function (PDF) of the amplitude of the CGL solutions, which is 
found to be approximately described by a stretched exponential distribution, 
$P(|A|) \approx e^{-|A|^\eta}$, where $\eta < 1$. This non-Gaussian PDF
is explained by the nonlinear characteristics of individual bursts combined
with the statistics of bursts.
Our results suggest a general picture in which an incoherent background of
weakly interacting waves, occasionally, `by chance', initiates intense, 
coherent, self-reinforcing, highly nonlinear events. 
\end{abstract}
\maketitle 

\section{Introduction}
  Many spatio-temporal dynamical systems show rare intense events. 
One example is that of large height rogue ocean waves \cite{Osborne1}.
Another example occurs in recent experiments on parametrically forced surface 
waves on water in which high spikes (bursts) on the free surface form 
intermittently in space and time \cite{Lathrop}.
Other diverse physical examples also exist (\it e.g.\rm, tornados, large
earthquakes, etc.). The characteristic feature of rare intense events 
is an enhanced tail in the event size probability distribution function. Here,
by enhanced we mean that the event size probability distribution function 
approached zero with increasing event size much more slowly than is the case 
for a Gaussian distribution. Thus these events, although rare, can be much
more common than an expectation based on Gaussian statistics would indicate.
The central limit theorem implies Gaussian behavior for a quantity that results
from the linear superposition of many random independent contributions. 
Non-Gaussian tail behavior can result from strong nonlinearity of the events, 
and enhancement of the event size tail might be expected if 
large amplitudes are nonlinearly self-reinforcing.  
Such nonlinear self-reinforcements is present in the nonlinear
Schr\"{o}dinger (NLS) equation. 
In particular, the two dimensional NLS equation, 
\begin{equation}
\frac{\partial A}{\partial t} = - i \alpha|A|^{2} A -i \beta
\bigtriangledown ^2 A.
\label{eq:nls}
\end{equation}
exhibits ``collapse" when the coefficients $\alpha$ and $\beta$ have the same
sign \cite{Bartuccelli}. 
In a collapsing NLS solution the complex field approaches 
infinity at some point in space, and this singularity occurs at finite time.
The NLS is conservative in that it can be derived from a Hamiltonian,
$\partial A / \partial t = - i \delta H / \delta A^* $, where $ H[A, A^*] = 
\frac{1}{2} \int [\alpha |A|^4 + \beta |\bigtriangledown A|^2] d {\bf x} $.
In the case of nonconservative dynamics, inclusion of lowest order dissipation 
and instability terms leads to the complex Ginzburg-Landau (CGL) equation
\cite{Levermore}. The CGL equation has been studied as a model for 
such diverse situation as chemical reaction \cite{Chemical}, 
Poiseuille flow \cite{Poiseuille}, 
Rayleigh-B\'{e}rnard convection \cite{Rayleigh}, and 
Taylor-Couette flow \cite{Taylor}. 
In the limit of zero dissipation/instability the CGL equation approaches the
NLS equation. For small nonzero dissipation/instability, the CGL equation
displays a solution similar to the NLS collapse solution, but with a large
finite (rather than infinite) amplitude at the collapse 
time \cite{Bartuccelli}. 
Furthermore, over a sufficiently large spatial domain, these events occur 
intermittently in space and time. Thus, in this parameter regime, the CGL 
equation may be considered as a model for the occurrence of rare intense events.

  In this paper we study the statistics and characteristics of rare intense
events in a two-dimensional CGL-based model. 
The probability distribution function (PDF) of the amplitude of 
the solutions is observed to be non-Gaussian in our 
numerical experiments. This non-Gaussian PDF is explained by the nonlinear
characteristics of individual bursts combined with the statistics of bursts.
The model equation we investigate is
\begin{eqnarray}
\frac{\partial A}{\partial t} &=& \pm A - (1+i \alpha)|A|^{2} A +
(1-i \beta) \bigtriangledown ^2 A \nonumber \\
                              & & +(\delta_r + i \delta_i) A^*,
\label{eq:cgl}
\end{eqnarray}
where $(\delta_r + i \delta_i) A^*$ is a parametric forcing term \cite{Park}.
We will consider two cases: one without parametric forcing ($\delta_r = 
\delta_i = 0$) in which case the plus sign is chosen in front of the first 
term on the right-hand side of (\ref{eq:cgl}) 
(Eq. (\ref{eq:cgl}) is then the usual CGL equation), 
and one with parametric forcing, in which case the minus sign is chosen.
As previously discussed, we choose our parameters so that our model,
Eq. (\ref{eq:cgl}), is formally close to the NLS equation (\ref{eq:nls}).
That is, we take $\alpha, \beta \gg 1, \delta_r, \delta_i$, and for our 
numerical solutions we will restrict attention to the case $\alpha = \beta$.
Note that the coefficient $\pm 1$ for the first term, as well as the ones in
$(1+i \alpha)$ and $(1 + i \beta)$ represent no loss of generality, since
these can be obtained by suitable scaling of the time($t$), the dependent
variable($A$), and the spatial variable(${\bf x}$).
  In Section II, we discuss the amplitude statistics of our two-dimensional
CGL models with and without the parametric forcing term. We find that the 
PDFs are approximately described by a stretched exponential distribution,
$P(|A|) \approx \exp(-|A|^{\eta})$, where $\eta$ is less than 1. 
In Section III, we investigate the characteristics of individual bursts. We
compare our numerical CGL results with known collapse solutions of the NLS
equation. The maximum amplitude obtained by many bursts (or the ``event size"
statistics) is discussed in Section IV. Section V presents further discussion 
and conclusions. 

  Our results lead us to the following picture for the occurrence of rare
intense events in our system. Linear instability and nonlinear wave saturation
lead to an incoherent background of small amplitude waves. This background
is responsible for the observed \it small \rm $|A|$ Gaussian behavior of our
probability distribution functions. When, by chance, the weakly interacting
waves locally superpose to create conditions enabling nonlinear, coherent 
self-reinforce, a localized, collapse-like event is initiated. Collapse takes
over, promoting large, rapid growth and spiking of $A$. This is followed by
a burn-out phase in which the energy in rapidly dissipated due to the 
generation of small scale structure by the spike. We believe that elements of
the above general picture may be relevant to a variety of physical situations
where rare intense events occur (\it e.g.\rm, the parametrically driven water
wave experiments in Ref. \cite{Lathrop}).
\begin{figure}[t]
\epsfig{file=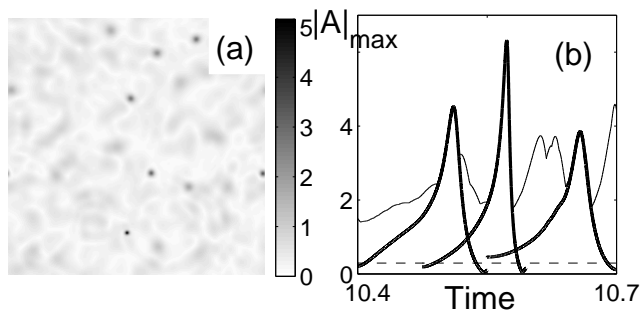,width= 8.5 cm}
\caption{Solutions of the CGL model. 
(a) Snapshot of the amplitude $|A|$ for $L = 20 \pi$, $\alpha = \beta = 30$, 
$\Delta t = 10^{-5}, \delta_r = \delta_i = 0$ and a $256 \times 256$ grid.
(b) Amplitude profile versus time. Solid line indicates $|A|_{max}$ of
the whole system, while thick solid lines indicate maximum amplitude of the 
localized events("bursts"). The dashed line indicates the average amplitude
of $|A|$ over the whole system $|A|_{avg} \sim 0.3$.}
\label{fig:pdf_11}
\end{figure}

\section{Amplitude Statistics}

\subsection{2D model without a parametric force ($\delta_r = \delta_i = 0$)}

  We first consider Eq.(\ref{eq:cgl}) with $\delta_r = \delta_i = 0$ and 
with the plus sign in the first term on the right hand side of the
equation. Figure \ref{fig:pdf_11}(a) shows a representation of $|A({\bf x},t)|$
[from numerical solution of Eq. (\ref{eq:cgl})] at a fixed instant $t$, where
large values are indicated by darker grey shades. As a function of time,
the localized dark shades occur in an seemingly random manner, become darker
(\it i.e., \rm increase their amplitude) and then go away (become light). 
Furthermore, the maximum amplitudes also display apparent randomness.
[see Fig. \ref{fig:pdf_11}(b)].
As shown in the next section (Sec. III), although the occurrence of these
intense events is apparently erratic in time and
space, individually these events are highly 
coherent. In this section, we will study the statistical properties of 
$A({\bf x},t)$.

  Our numerical solutions of (\ref{eq:cgl}) employ periodic boundary 
conditions on a $256 \times 256$ grid. 
We choose the parameters, $\alpha$ and $\beta$, large enough 
($\alpha = \beta = 30$) so that the solutions of our model are close to 
solutions of the NLS equation. We choose the time step 
small enough to satisfy the condition for unconditional stability 
of our second-order accurate time integration($\Delta t = 10^{-5}$). 
We use random initial condition (at $t=0$, we generate random values for
amplitudes and phases at each grid point).
Localized structures, "bursts", develop very rapidly and occur throughout the
spatial domain. The typical life time of a burst ($\delta t$) is approximately 
0.2 time units. The maximum amplitudes of bursts are different for different 
burst events. 

  Imagining that we choose a space-time point $({\bf x},t)$ at random, we now
consider the probability distribution functions for $|A|$ (the magnitude of
$A$), $A_r = Re[A]$ (the real part of $A$), and $A_i = Im[A]$ (the imaginary
part of $A$). We denote these distribution functions $P(|A|), P_r(A_r),
P_i(A_i)$, and we compute them via histogram approximations using the values
of $|A|, A_r,$ and $A_i$ from each of the $256 \times 256$ grid points at many 
time frames \cite{footnote}. We find that these distributions are independent
of the periodicity length $L$ used in the computation as long as it is
sufficiently large compared to the spatial size of a burst, but is not so large
that spatial resolution on our $256 \times 256$ grid becomes problematic.
In our computations of $P, P_r$ and $P_i$, we choose $L = 20 \pi$.

\begin{figure}[t]
\epsfig{file=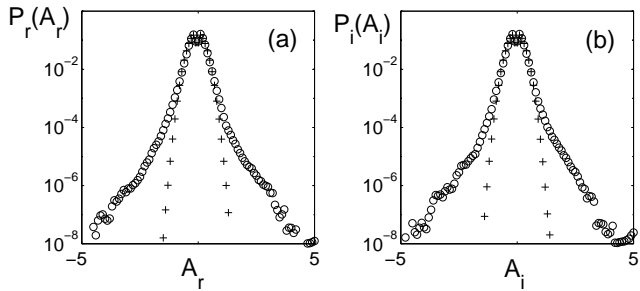,width= 8.5 cm}
\caption{Probability distribution functions obtained from numerical solution
of Eq.(2) using the same parameters as in Fig. 1. The circles are data for
$P_r$ and $P_i$, while the pluses are the probability distributions $P_r'$
and $P_i'$ obtained from the phase randomized amplitude.}
\label{fig:pdf_21}
\end{figure}

  Figures 2 show the numerically computed probability distributions $P_r(A_r)$
[Fig. 2(a)] and $P_i(A_i)$ [Fig. 2(b)] plotted as open circles. Since 
Eq. (\ref{eq:cgl}) with $\delta_r = \delta_i = 0$ is invariant to the 
transformation 
$ A \rightarrow A \exp(i\phi)$ (where $\phi$ is an arbitrary constant), we 
expect the distribution $P_r$ and $P_i$ to be the same to within the 
statistical accuracy of their determinations. This is born out by Figs. 2.
In order to highlight the essential role that nonlinearity plays in 
determining these distribution functions, we have also computed them after 
randomizing the phases of each Fourier component. That is, representing
$A({\bf x},t)$ as
\begin{equation}
A({\bf x},t) = \sum_{{\bf k}} a_{{\bf k}}(t) \exp (i {\bf k}\cdot{\bf r}),
\end{equation}
where ${\bf k} = (2m\pi/L, 2n\pi/L)$, we form a new amplitude,
$A'({\bf x},t)$ as
\begin{equation}
A'({\bf x},t) = \sum_{{\bf k}} a_{{\bf k}}(t) 
\exp (i {\bf k}\cdot{\bf r} + i \theta_{{\bf k}}),
\end{equation}
where for each ${\bf k}$, the angle $\theta_{{\bf k}}$ is chosen randomly
with uniform probability density in $[0, 2\pi]$. The probability distribution
functions for the real and imaginary parts of the randomized amplitudes 
$A'$ are shown as pluses in Figs. \ref{fig:pdf_21}. 
Note that by construction, $A$ and $A'$ have
identical wavenumber power spectra. Due to the random phases, $A'$ at any
given point $\bf{x}$ can be viewed as a sum of many independent random numbers
(the Fourier components). Hence the $P_r$ and $P_i$ distributions are 
expected to be Gaussian, $\log P_{r,i} \sim [(const.) - (const.) A_{r,i}^2]$. 
This is confirmed by the semi-log plots of Figs. 2, where the data plotted as 
pluses can be well-fit by parabolae.

  The above comparisons with the phase randomized variable
$A'$ are motivated by imagining the hypothetical situation where the amplitude 
is formed by the superposition of many noninteracting linear plane waves.
In that case we would have an amplitude field of form
\begin{equation}
\sum_{{\bf k}} b_{{\bf k}}(t) \exp (i {\bf k}\cdot{\bf r}+i \omega_{{\bf k}}t).
\end{equation}
Because $\omega_{{\bf k}}$ is different for different ${\bf k}$, the phases 
become uncorrelated for sufficiently large time $t$.

\begin{figure}[t]
\epsfig{file=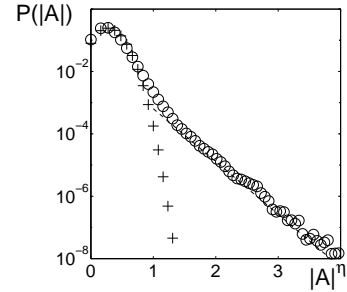,width= 4.5 cm}
\caption{Probability distribution functions before randomizing the phases of
the solutions (o) and after randomizing the phases (+). Note the horizontal
axis is $|A|^\eta$, where the exponent $\eta=0.8$ is chosen to yield 
approximately linear dependence of $\log[P(|A|)]$ on $|A|^\eta$ for large
values of $|A|$.}
\label{fig:pdf_31}
\end{figure}

  Comparing the data from $A$ and $A'$ in Figs. 2, for small values of $A_r$
and $A_i$, the PDFs are nearly Gaussian. This can be attributed to near linear
behavior of small amplitude waves. On the other hand, for the tails of the
distributions, we note substantial enhancement relative to the Gaussian 
distributions. These must be due to coherent phase correlations resulting from
nonlinear interaction of different wavenumber components of $A$. Such phase
coherence is implied by the observed coherent localized burst structures. 

  Figure 3 shows the numerically obtained distribution $P(|A|)$ plotted as 
circles and the probability distribution for the phase randomized amplitude
$|A'|$ plotted as pluses. 
Again, the enhancement of the large amplitude tail
is evident. Note that the vertical axis in Fig. 3 is logarithmic, while the
horizontal axis is $|A|^{\eta}$. Here we choose the power $\eta = 0.8$ so that
the large $|A|$ data in this plot are most nearly fit by a straight line. That
is, we attempt to fit $P(|A|)$ using a stretched exponential. The slope of the 
dashed straight line in the figure is chosen to match the large $|A|$ data.
Thus, over the range of $|A|$ accessible to over numerical experiment, we
find that the enhanced large $|A|$ tail probability density is roughly fit by
a stretched exponential,
\begin{equation}
P(|A|) \sim \exp(-\zeta |A|^{0.8}).
\end{equation}

\subsection{2D model with parametric forcing ($\delta_r,\delta_i \ne 0$)}

  We now report similar results for the case of parametric forcing, Eq.
(\ref{eq:cgl}) with $\delta_r, \delta_i \ne 0$ and the minus sign chosen
in the first term on the right side of (\ref{eq:cgl}). In this case, instability
of small amplitude waves is caused by the parametric forcing (nonzero
$\delta_{r,i}$) and the $-A$ term represents a linear wavenumber independent
damping. This model for parametric forcing was introduced \cite{Corellet}
and has been used to model various situations. One such situation is that of
periodically forced chemical reactions \cite{Hagberg}. Our motivation for
considering this model is the Faraday experiments on strong parametric forcing
of surface water waves in Ref. \cite{Lathrop}. In that work intermittent
formation of large localized surface perturbations results in splash and 
droplet formation.

\begin{figure}[t]
\epsfig{file=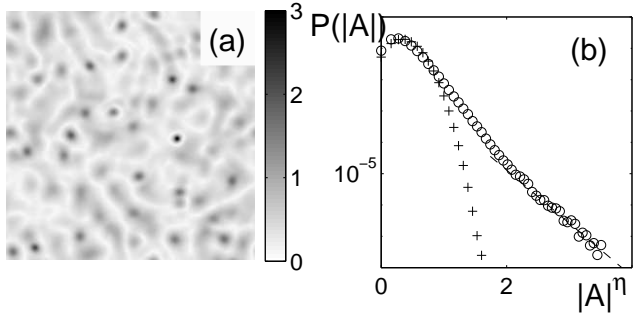,width= 8.5 cm}
\caption{Solutions of the model with parametric forcing. 
(a) snapshot of $|A|$. Dark regions have high amplitudes.
(b) $P(|A|)$ versus $|A|^\eta$, where $\eta = 0.8$. (See captions in Fig.
\ref{fig:pdf_31}.)} 
\label{fig:pdf_12}
\end{figure}

  Parameters for our numerical simulations are the same as in Sec. IIA except
that now $\delta_r = \delta_i = 5$. Again coherent localized structures, 
"bursts", develop rapidly and occur intermittently throughout the spatial 
domain, Fig. \ref{fig:pdf_12}(a). As in Sec. IIA, the typical life 
time of a burst is less than 0.2 time units, and the maximum amplitudes of 
bursts are different for different bursts. 

\begin{figure}[b]
\epsfig{file=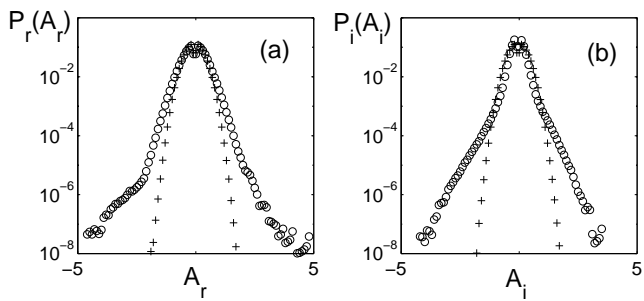,width= 8.5 cm}
\caption{$P_r(A_r)$ and $P_i(A_i)$ with parametric forcing. The numerical
parameters are the same as those for Fig. \ref{fig:pdf_12}. 
The circles are PDFs for $A_r$ and $A_i$ before randomizing the phases of the 
solutions, while pluses are PDFs after randomizing the phases.}
\label{fig:pdf_22}
\end{figure}

  The PDF, $P(|A|)$ again shows a stretched exponential tail with exponent
$\eta = 0.8$, Fig. \ref{fig:pdf_12}(b).
The circles in Figs. \ref{fig:pdf_22} show the PDFs of the real and imaginary 
parts of $A$, while the pluses are data for the PDFs after randomizing the 
phase. The shape of the PDFs around the central part are nearly Gaussian.
In contrast, at large amplitude the PDFs are significantly non-Gaussian.
A major difference with the case $\delta_r = \delta_i = 0$ is that, 
with parametric forcing, the model is not invariant to 
$A \rightarrow A e^{i\phi}$, and thus $P_r$ and $P_i$ may be expected to 
evidence differences not present for $\delta_r = \delta_i = 0$. 
This is seen to be the case in Figs. \ref{fig:pdf_22}.  

\section{Characteristics of Individual Burst Events}

  Solutions of the CGL equation with large $\alpha$ and $\beta$ may be expected
to have features in common with solutions of the NLS equation. 
It is known that the NLS equation yields localized events which develop finite
time singularities where the amplitude becomes infinite ~\cite{Levermore}. 
While it is difficult to understand the dynamics of the solutions of CGL 
equation from direct rigorous analysis, the solutions of the 
NLS equation are relatively well understood. Thus, we analyze the dynamics of
individual CGL bursts guided by the known localized self-similar collapsing 
solution of the NLS equation. 

  The NLS equation has a special solution \cite{Fibich} of the form
\begin{equation}
A = e^{i\theta t}R(r), ~~~~r=\sqrt{x^2+y^2},
\end{equation}
where the radial function $R(r)$ satisfies
\begin{eqnarray}
\label{eq:radial}
\left( \frac{\partial ^2}{\partial r^2} + \frac{1}{r} 
\frac{\partial}{\partial r} \right)R - \xi R + R^3 = 0, \\ \nonumber 
~~ \left| \frac{\partial R}{\partial r} \right|_{r=0} = 0, ~~R(\infty) = 0,
\end{eqnarray}
where $ \alpha = \beta$, $\xi = \theta / \beta$ .
Since (\ref{eq:nls}) is invariant under the scaling transformation 
\cite{Lemesurier}, 
\begin{equation}
A({\bf x},t) \longrightarrow L(t)^{-1} A({\bf \kappa}, \tau)
{\rm e}^{iL \dot{L}|{\bf \kappa}|^2/4},
\label{eq:rescalea}
\end{equation}
where $L(t)$ tends to zero as $t^* \longrightarrow t$, $t<t^*$ and
\begin{equation}
{\bf \kappa} = \frac{{\bf x}}{L(t)}, ~~~~ 
\tau = \int_0^t\frac{1}{L^2(s)}ds,
\label{eq:rescalex}
\end{equation}
a family of collapsing solutions of the NLS is given by the rescaled solution
of Eq. (\ref{eq:radial}). 

\begin{figure}[t]
\epsfig{file=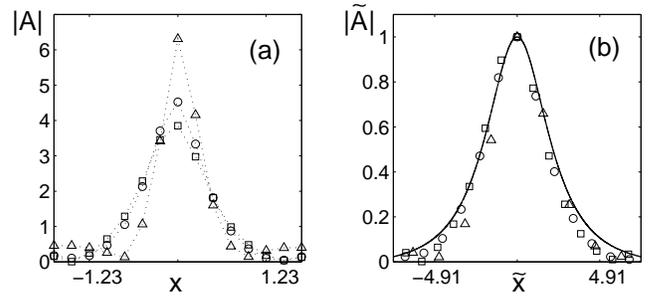,width= 8.5 cm}
\caption{Self-similar bursts. (a) Enlarged plots of a burst at 
$t_1 (\bigcirc) = 10.448, $ at grid point $(x,y) = (15.2,22.1),$ and 
$t_2 (\triangle) = 10.530,$ at grid point $(51.1,25.5),$ and 
$t_3 (\square)= 10.644 $ at grid point $(57.4,38.8)$.
(b) Scaled profiles, where $|\tilde{A}|= |A|/L$, $\tilde{x} = x/L$, 
and $L = |A|_{max}$ at $t_1$, $t_2$, and $t_3$. The solid line represents 
the radial solution of Eq. (\ref{eq:radial}).}
\label{fig:bursts}
\end{figure}

With these considerations, we test the expected approximate self-similarity of 
individual bursts observed in our numerical solutions of Eq. (\ref{eq:cgl}).
We consider three typical bursts that occur at different times and spatial 
positions. In particular, we choose these three as the three dark regions
in Fig. \ref{fig:pdf_11}(a) whose spatial maxima have the time dependences
shown as thick solid lines in Fig. \ref{fig:pdf_11}(b).

  In Fig. \ref{fig:bursts}(a) we plot the $x$-dependence of $|A|$ at constant
$y$ for each of these bursts at the time that they reach their maximum 
amplitude (the positions in $x$ of the maxima have been shifted to $x=0$ 
and the constant $y$ value for each is at the location of $|A|_{max}$). 
Note that, when they reach their maxima, the three bursts have different
amplitudes and width. We rescale the amplitude and spatial coordinate as
suggested by (\ref{eq:rescalea}) and (\ref{eq:rescalex}), 
$|\tilde{A}|=|A|/L$ and $\tilde{x}=x/L$, and we take $L = |A|_{max}$ (which
normalizes $|\tilde{A}|_{max}$ to one).
The resulting data are plotted in Fig. \ref{fig:bursts}(b) along with the 
solution of Eq. (\ref{eq:radial}).
[We again rescale $R(r)$ using (\ref{eq:rescalea}) and (\ref{eq:rescalex}), 
and we note that, after this rescaling, the result is independent of the value 
of $\xi=0.1$ in Eq. (\ref{eq:radial}).] The three burst profiles show evidence 
of collapsing onto the theoretical radial solution. 

\begin{figure}[t]
\epsfig{file=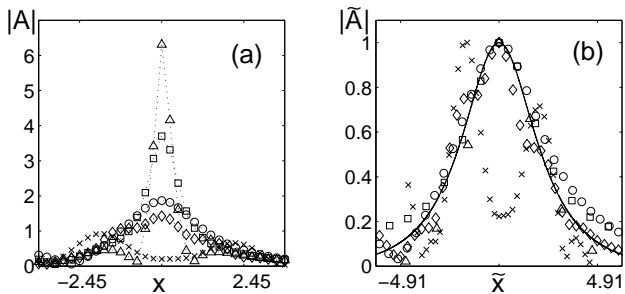,width= 8.5 cm}
\caption{Self-similarity of single burst. (a) Enlarged plots of a burst at 
$t_1 (\bigcirc) = 10.50$, $t_2 (\square) = 10.52$, $t_3 (\triangle)= 10.53$,
$t_4 (\diamond) = 10.54$, and $t_5 (\times) = 10.55$. 
(b) Scaled profiles, where $|\tilde{A}|= |A|/L$, $\tilde{x} = x/L$, 
and $L = |A(r)|_{max}$ at $t_1, t_2, t_3 $, and $t_4$. The solid line 
represents the solution of Eq. (\ref{eq:radial}).}
\label{fig:single1}
\end{figure}

  Now, we consider the time dependence of a single burst. We select the burst 
at the grid point $(x,y) = (51.1,25.5)$ (see caption to Fig. \ref{fig:bursts})
and investigate the evolution of its shape and height. 
We display profiles of the burst at 5 different times in 
Fig. \ref{fig:single1}(a). Rescaling each profile using Eq. (\ref{eq:rescalea})
and Eq. (\ref{eq:rescalex}), and defining $L$ in the same way as before,
the four profiles at the first four times approximately collapse onto the 
radial solution of (\ref{eq:radial}) as shown in Fig. \ref{fig:single1}(b). 
When the burst reaches its maximum amplitude, the amplitude at some distance 
away from the center becomes zero (see the amplitude profile at $t=t_3$). 
After that, the center decays very rapidly 
(see the amplitude profile at $t=t_5$). 
(Note that in this section and the next section, we present numerical results 
for (\ref{eq:cgl}) with $\delta_r =\delta_i = 0$. However, we have also 
verified that the CGL model with parametric forcing also has similar self 
similarity properties.)

\section{Statistics of Bursts}

  The self similar properties of bursts implies that the solutions of the 
CGL model consist of self-similar bursts of various maximum heights. 
Thus, we expect that the enhanced tail 
(the deviation from the Gaussian distribution) 
of the amplitude probability distribution $P(|A|)$ can be understood by 
the statistics of bursts. In particular, we consider $g(h)$, the frequency of 
bursts which have maximum height $h$, and a distribution $P_j(|A|)$ defined 
for an individual burst (burst $j$), as follows.

\begin{figure}[t]
\epsfig{file=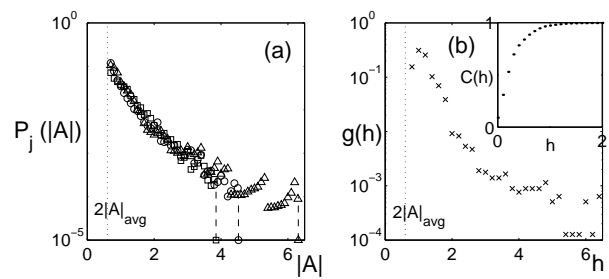,width= 8 cm}
\caption{Statistics of localized events("bursts"). 
(a) $P_{h_j}(|A|)$ at three different times: 
$t_1 (\bigcirc) = 10.448$ and $h=4.52$, $t_2 (\triangle) = 89.0$ and 
$h=6.31$, and $t_3 (\square) = 94.0$ and $h=3.85$. 
(b) The frequency of bursts which have maximum height $h$, $g(h)$. The inset
indicates $C(h)$ versus $h$ defined in Eq. (\ref{eq:normal}).}
\label{fig:stat}
\end{figure}

  We define the time interval for each burst as the time between when its peak 
value exceeds $2|A|_{avg}$ and when its peak value drops below $2|A|_{avg}$ 
[typically the time duration of a burst is less than $0.2$, see 
Fig. \ref{fig:pdf_11}(b)]. Here $|A|_{avg}$ is the space average of $|A|$ over 
the entire spatial grid of the simulation at each time $t$ [$|A|_{avg}$ is 
approximately constant at about $0.3$ over all time steps in the simulation, 
see the dashed line in Fig. \ref{fig:pdf_11}(b)]. Consistent with the 
observation that a typical burst has radial symmetry, we define the domain of 
the burst to be a circular region of radius $r_{eff}$ centered at the burst 
maximum, where $r_{eff}$ is the maximum radius of a circle such that the 
average of $|A|$ over the perimeter of the circle is greater than $2|A|_{avg}$ 
(typically, $1.23 \le r_{eff} \le 4.91$).
In Fig. \ref{fig:stat}(a) we show the distribution $P_j(|A|)$ for the three
bursts in Fig. \ref{fig:pdf_11}(b) (thick solid lines). The first burst 
($j=1$) has $h = 4.52$ and is plotted as the open circles in
Fig. \ref{fig:stat}(a); the second burst ($j=2$) has $h = 6.31$ and is plotted
as the open triangles; and the third burst ($j=3$) has $h = 3.85$ and is 
plotted as the open squares. 
These distributions are obtained from histograms of the 
values of $|A|$ at grid points in the domains and time steps in the duration 
of each of these bursts. 
We obtain $g(h)$ by counting the number of bursts which have maximum heights 
between $h$ and $h + \Delta h$, where $\Delta h = 0.2$ 
[see Fig. \ref{fig:stat}(b)]. (In Fig. \ref{fig:stat}(a) $P_j(|A|)$ is not
plotted for $|A| < 2|A|_{avg}$, since, by our procedure this range lacks 
meaning, and since we are interested in the behavior at large values of $|A|$.)
We note that the $P_j(|A|)$ in Fig. \ref{fig:stat} all approximately coincide 
for $|A|<h$. Thus the only characteristic of the bursts on which $P_j(|A|)$
depends is the maximum burst amplitude $h$ at which $P_j(|A|)$ goes to zero.
To incorporate this fact, we replace $P_j(|A|)$ by the notation $P_h(|A|)$. 

  The above suggests that $P(|A|)$ can be obtained from the
following approximation \cite{Iwasaki}
\begin{equation}
P(|A|) \sim \int_0^{\infty} g(h)P_h(|A|)dh,
\label{eq:approx}
\end{equation}
where $P_h(|A|)$ is a probability distribution of a single burst whose temporal
maximum amplitude is $h$. Since $P_h(|A|)$ vanishes for $ |A| > h$ and 
because the $P_h(|A|)$ approximately coincide for $|A|<h$ 
[see Fig. \ref{fig:stat}(a)], we approximate $P_h(|A|)$ as
\begin{eqnarray}
P_h(|A|) &\sim& C^{-1}(h) \theta (h-|A|) P_*(|A|), \\
\label{eq:ph}
C(h) &=& \int_0^h P_*(|A|)d|A|, 
\label{eq:normal}
\end{eqnarray}
where $C(h)$ is a normalization factor [$C(h) \sim 1$, when $h>1$; see
the inset on Fig. \ref{fig:stat}(b)], 
$\theta(h-|A|)$ is a step function, and $P_*$ is the distribution that we
numerically compute at the largest value of $h$ that we considered 
($h_{max}=6.31$). Using (\ref{eq:approx}) and (\ref{eq:ph}), we can further 
approximate $P(|A|)$ as
\begin{eqnarray}
P(|A|) &\sim& \int_0^{\infty} C^{-1}(h) 
         \theta(h-|A|)g(h)P_*(|A|)dh \nonumber \\
       &\sim& P_*(|A|) \int_{|A|}^{\infty} C^{-1}(h) g(h) dh.
\label{eq:gh}
\end{eqnarray}
(The integral in (\ref{eq:gh}) is the cumulative frequency of bursts which 
have maximum height greater than $|A|$.)

\begin{figure}[t]
\epsfig{file=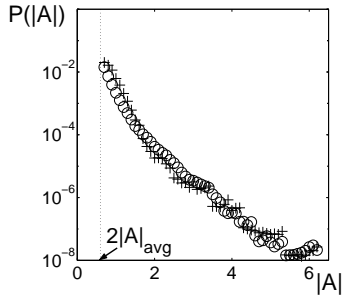,width= 4.5 cm}
\caption{$P(|A|)$ versus $|A|$. Circles represent $P(|A|)$ obtained directly
from our numerical solutions of (\ref{eq:cgl}), while pluses represent
$P(|A|)$ obtained using $P_*(|A|)$ and $g(h)$ from Fig. \ref{fig:stat},
and Eqs. (\ref{eq:approx})-(\ref{eq:gh}).} 
\label{fig:tot}
\end{figure}

   Figures \ref{fig:stat} show the numerically obtained $g(h)$ and 
$P_*(|A|)$. Inserting $P_*(|A|)$ into Eq.(\ref{eq:normal}) and
Eq.(\ref{eq:gh}), we obtain the prediction for $P(|A|)$ plotted as pluses in
Fig. \ref{fig:tot} for $|A|>2|A|_{avg}$. This appears to agree well with the 
$P(|A|)$ obtained from our numerical solutions of (\ref{eq:cgl}) 
(open circles). (Note that we shift the predicted $P(|A|)$ (pluses) to the 
$P(|A|)$ (open circles) obtained from (\ref{eq:cgl}) after removing data points 
for $|A|<2|A|_{avg}$.)
 
\section{Conclusion}

  We find that the large $A$ behavior of the PDF obtained from our CGL 
solutions is approximately described by a stretched exponential form, 
$P(|A|) \approx e^{-|A|^\eta}$, where $\eta < 1$. 
In addition, for small $A$, $P_r(A_r)$ and $P_i(A_i)$ are approximately
Gaussian, as is the case for a random linear superposition of waves.
We also observe the self similar properties of individual bursts, which allow 
us to consider the large amplitude behavior of our CGL solutions as composites 
of coherent self similar bursts. Based on this we explain the observed 
non-Gaussian $P(|A|)$ using the nonlinear characteristics of individual 
bursts $P_h(|A|)$ combined with the statistics of burst occurrences $g(h)$.

  These results lead us to conjecture the following picture of rare intense
events in our model. Linear instability leads to a background of relatively
low amplitude waves that are weakly interacting and result in a 
random-like, incoherent background and low $|A|$ Gaussian behavior of
$P_r(A_r)$ and $P_i(A_i)$. When, by chance, this incoherent behavior results
in local conditions conductive to burst formation, nonlinear, coherent, 
self-reinforcing collapse takes over and promotes a large growth and spiking
of $A$. We believe that this general mechanism may be operative in a variety
of physical situations in which rare intense events occur (\it e.g.\rm,
the water wave experiments of Ref. \cite{Lathrop}).

  We thank D. P. Lathrop for initial discussion and for attracting our 
attention to the subject of rare intense events. We thank P. N. Guzdar for his
advice on numerics. This work was supported by the Office of Naval Research 
(Physics) and by the National Science Foundation (PHY0098632).


\begin{references} 
\bibitem{Osborne1} A. R. Osborne, M. Onorato, and M. Serio, Phys. Lett. A 
\bf 275 \rm, 386(2000).
\bibitem{Lathrop} J. E. Hogrefe, \it et al\rm., Physica D 
\bf 123 \rm 183(1998); B. W. Zeff, \it et al\rm., Nature \bf 403 \rm 401(2000).
\bibitem{Bartuccelli} M. Bartuccelli, \it et al.\rm, Physica D \bf 44 \rm, 
421(1990).  They draw a phase diagram figures for NLS equation (Fig. 1) and 
CGL equation (Fig. 2).
\bibitem{Levermore} C. D. Levermore and M. Oliver, Lect. Appl. Math. \bf 31 \rm
141(1996). 
\bibitem{Chemical} Y. Kuramoto, \it Chemical Oscillations, Waves and 
Turbulence \rm, Series in Synergetics \bf 19 \rm, Springer, New York, 1984.
\bibitem{Poiseuille} K. Stewartson and J.T. Stuart, J. Fluid Mech. \bf 48 
\rm, 529(1971).
\bibitem{Rayleigh} A.C. Newell and J.A. Whitehead, J. Fluid Mech. 
\bf 38 \rm, 279(1969).
\bibitem{Taylor} G. Ahlers and D.S. Cannell, Phys. Rev. Lett. \bf 50 \rm, 
1583(1983).
\bibitem{Park} H.-K. Park, Phys. Rev. Lett. \bf 86 \rm 1130(2000); also see
the references therein.
\bibitem{footnote} $P(|A|)$ is to be contrasted with the probability 
distribution of a single burst amplitude $P_h(|A|)$ to be discussed 
in Sec. IV.
\bibitem{Corellet} P. Corellet, Phys. Rev. Lett \bf 56 \rm, 724(1986).
\bibitem{Hagberg} C. Elphick, A. Hagberg, and E. Meron, Phys. Rev. Lett 
\bf 80 \rm, 5007(1998).
\bibitem{Fibich} G. Fibich and D. Levy, Phys. Lett. A \bf 249 \rm, 286(1998).
\bibitem{Lemesurier} B. LeMesurier, \it et al.\rm, Physica D 
\bf 32 \rm, 210(1988).
\bibitem{Iwasaki}  H. Iwasaki and S. Toh, Prog. Theo. Phys. \bf 87\rm, 1127 
(1992).
\end{references}
\end{document}